\documentstyle{mn}
\title{Acceleration time scale in an ultrarelativistic shock}

\author[Janusz Bednarz]
       {Janusz Bednarz \\
Obserwatorium Astronomiczne, Uniwersytet Jagiello\'nski, ul. Orla 171,
30-244 Krak\'ow, Poland}

\begin{document}

\maketitle

\label{firstpage}

\begin{abstract}
The acceleration mechanism at ultrarelativistic shocks is investigated
using the Monte Carlo simulations. We apply a method of discrete small
amplitude particle momentum scattering to reproduce highly anisotropic
conditions at the shock and carefully describe the acceleration mechanism.
The obtained acceleration times equal $1.0\, r_{g}/c$ if the spectral index
reach the value of $2.2$, independent of physical conditions in the shock.
Some other parameters of the acceleration process are also provided.
\end{abstract}


\section{Introduction}

Observations carried out by the Burst and Transient Source Experiment show
that GRBs originate from cosmological sources (Meegan et al. 1992 and Dermer
1992). Identification of the host galaxy for the GRB 971214 (Kulkarni et al.
1998) and several other bursts causes there is little doubt now that some,
and most likely all GRBs
are cosmological. These phenomena are surely related to ultrarelativistic
shocks with the Lorentz factors $\gamma>10^{2}$.

Several papers suggested
that ultrarelativistic shocks in GRBs could be sources of high energy
cosmic rays (cf. Waxman 1995, Vietri 1995), and simulations done by Bednarz
\& Ostrowski (1998) showed that such shocks are able to accelerate charged
particles and values of their energy spectral indices converge to
$\sigma=2.2$ when $\gamma \rightarrow \infty$ and/or magnetic turbulence
amplitudes grow. Because the acceleration mechanism is quite different from
that in the non-relativistic and mildly relativistic regime we distinguish
a class of ultrarelativistic shocks if their Lorentz factors $\gamma\gg 1$.

Observations seem to confirm this mechanism. Waxman (1997) used a fireball
model of GRBs and showed from the functional dependence of the flux on time
and frequency that $\sigma=2.3\pm0.1$ in the afterglow of GRB 970228.
Galama et al. (1998) made two independent measurements of the electron
spectrum index in the afterglow of GRB 970508 which was very close to $2.2$.

\section{Acceleration mechanism}

A particle crossing the shock to upstream medium has a momentum vector
nearly parallel to the shock normal. Then the particle momentum changes
its inclination in two ways by: 1) scattering in an inhomogeneous magnetic
field and 2) smooth variation in a homogeneous field component. Hereafter,
the mean deflection angle in these two cases will be denoted by
$\Delta \Omega_{S}$ and $\Delta \Omega_{H}$, respectively. The first
process is a diffusive one and the second depends on time linearly. That
means that with increasing shock velocity, keeping other parameters
constant, $\Delta \Omega_{S}$ decreases slower as a square root of time
in comparison with $\Delta \Omega_{H}$. The Lorentz transformation shows
that with $\gamma\gg 1$ even a tiny angular deviation in the upstream
plasma rest frame can lead to a large angular deviation in the downstream
plasma rest frame. Let us denote a particle phase by $\phi$ and the
angle between momentum and a magnetic field vector by $\theta$ both
measured in the downstream plasma rest frame. Values of these parameters
at the moment when a particle crosses the shock downstream determine if it
is able to reach the shock again in the case of neglected magnetic field
fluctuations downstream of the shock. In fact a motion in the homogeneous
magnetic field carries a particle in such a way that it cannot reach the
shock again. The magnetic field fluctuations upstream of the shock perturbing
the momentum direction lead to broadening the ($\phi,\theta$) range
that allows particles to reach the shock again. Thus, as we show below
for efficient scattering, when $\Delta \Omega_{H}$ becomes unimportant in
comparison to $\Delta \Omega_{S}$, the spectral index and the acceleration
time reach their asymptotic values.

The discussed relation between
$\Delta \Omega_{H}$ and $\Delta \Omega_{S}$ is reproduced in our
simulations and presented in Fig.~1. There are shown 11 points from
$\gamma$ = 100 to 320 and three additional for $\gamma$ = 640, 1280, 2560.
The expected linear dependence of these quantities can be noticed.

\begin{figure}
\vspace*{5.8cm}
\includegraphics{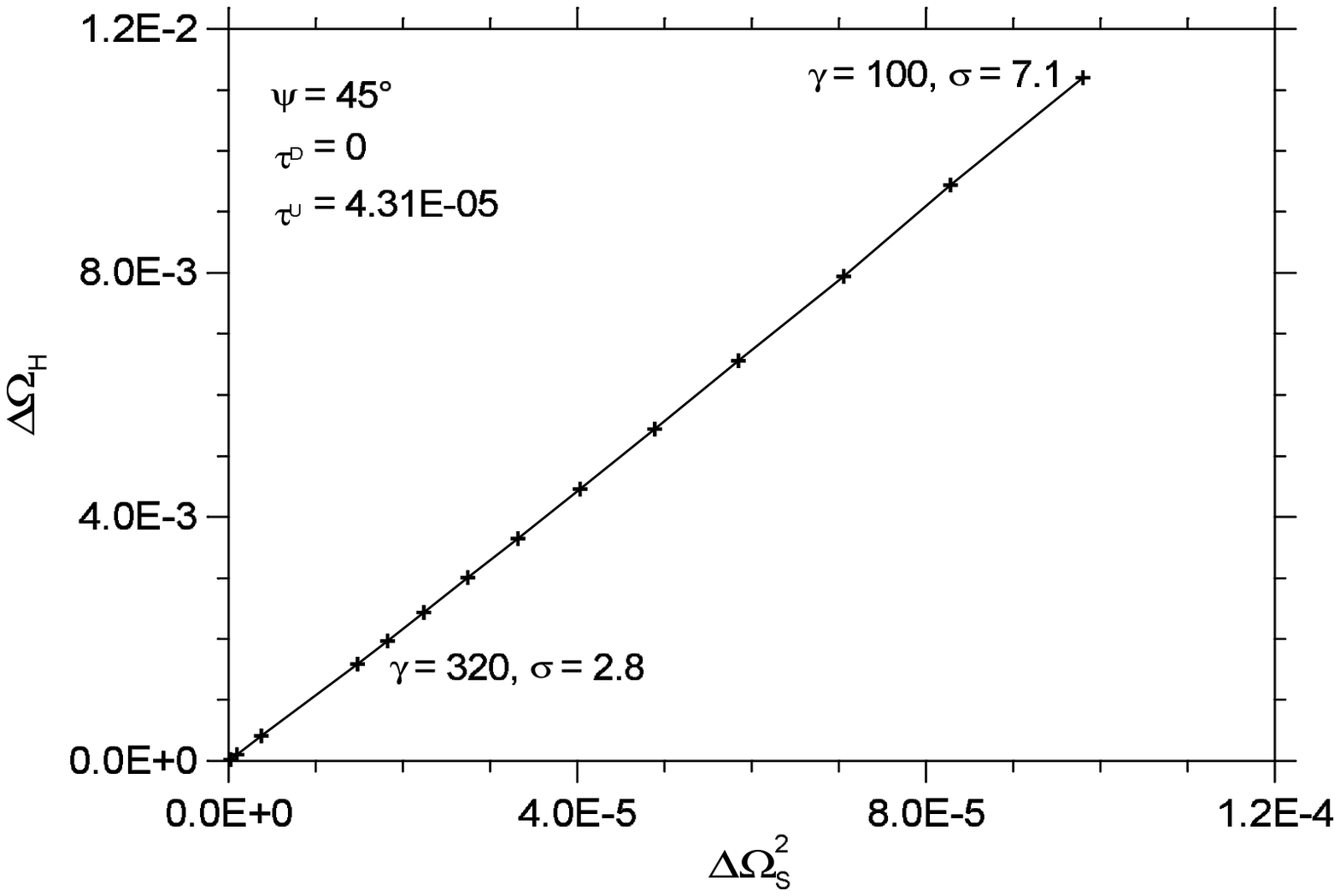}
\caption{The relation between the mean deflection angle upstream of
the shock caused by the scattering in an inhomogeneous
magnetic field ($\Delta \Omega_{S}$) and by smooth variation
in a homogeneous magnetic field ($\Delta \Omega_{H}$).
Last three points for $\Delta \Omega_{H}$ below $1\cdot 10^{-3}$
represent $\gamma$ = 640, 1280, 2560 and yield $\sigma$ =
2.5, 2.3 and 2.2 respectively.}
\label{fig1}
\end{figure}

\begin{figure}
\vspace*{5.8cm}
\includegraphics{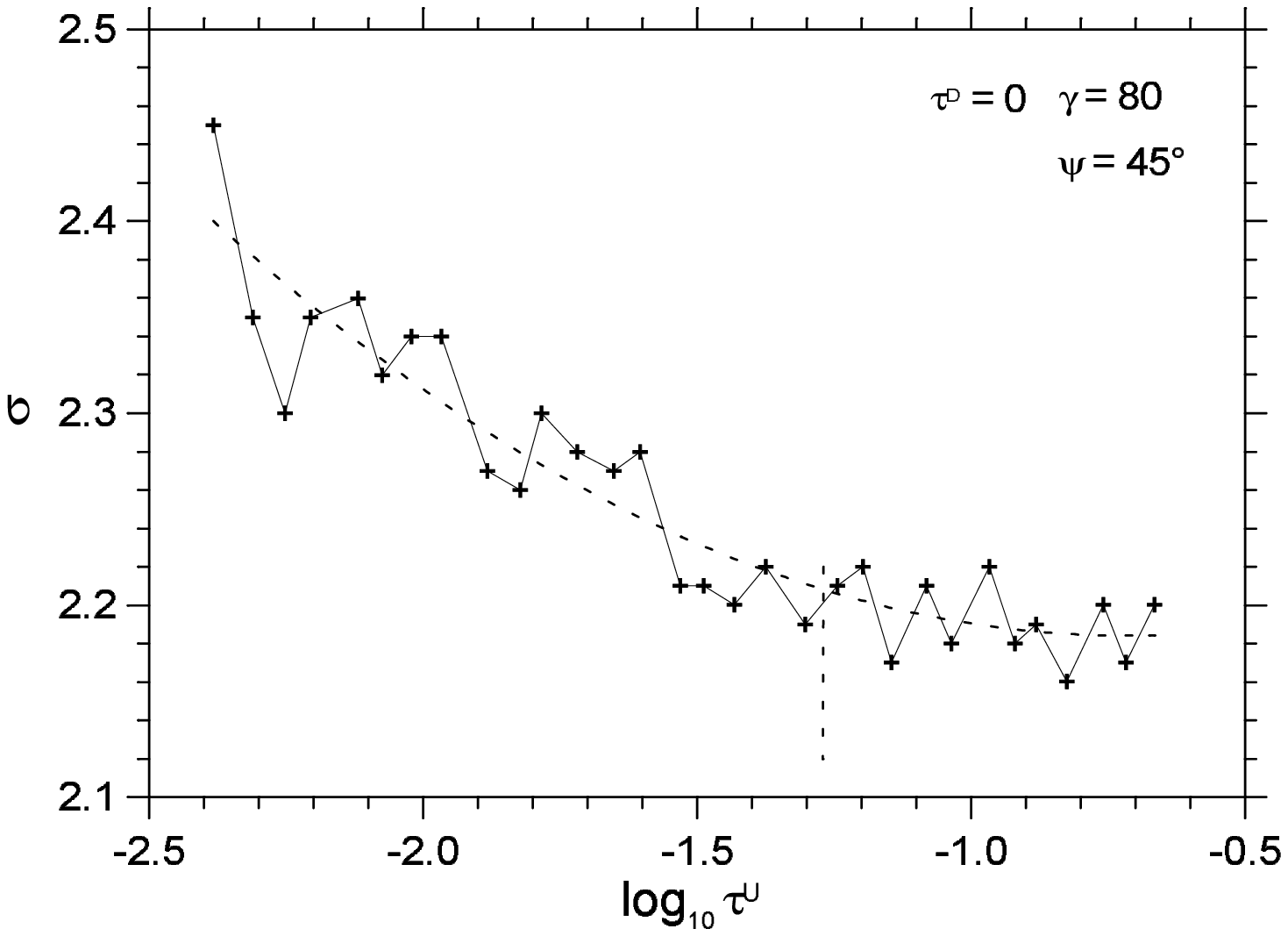}
\caption{
Simulated spectral indices as a function of magnetic field fluctuations
upstream of the shock. Fluctuations downstream of the shock are neglected.
The chosen $\tau^{U}$ value for $\gamma=80$ and $\psi=45^\circ$ is
pointed by a dashed line. A second-degree polynomial fit is also marked by
a dashed line.}
\label{fig2}
\end{figure}

\section{Numerical simulations}

In simulations we follow the procedure used by Bednarz \& Ostrowski (1996)
with a hybrid approach used in Bednarz \& Ostrowski (1998). Monoenergetic
seed particles are injected at the shock and then their trajectories are
derived in the perturbed magnetic field. The inhomogeneities are simulated
by small amplitude particle momentum scattering within a cone with angular
opening $\Delta \vartheta$ less than the particle anisotropy $\sim 1/\gamma$
(cf. Ostrowski 1991).

A particle is excluded from simulations if it escapes through the
free-escape boundary placed far off the shock or reaches the energy larger
than the assumed upper limit. These particles are replaced with the ones
arising from splitting the remaining high-weight particles, preserving their
physical parameters. Particles that exist longer than the time upper limit
for simulations are excluded from simulations without replacing.

All computations are performed in the respective upstream or downstream plasma
rest frame. When
particles cross the shock their parameters are transformed to the current
plasma rest frame and the weighted contribution divided by the particle
velocity component normal to the shock ($\equiv$ particle density) is added
to the time and momentum bin depending on particle parameters, as measured in
the shock normal rest frame. For the considered continuous injection
after initial time, the energy cut-off of the formed spectrum shifts toward
higher energies with time. The resulting spectra allows one to fit spectral
indices and derive acceleration time in the shock normal rest frame in units
of downstream $r_{g}/c$ ( $r_{g}$ - particle gyroradius in the homogeneous
magnetic field component, $c$ - speed of light; for details see Bednarz \&
Ostrowski 1996). We transform the acceleration time $t_{acc}$ to the
downstream plasma rest frame.
Hereafter, subscripts U or D mean that a parameter is measured in the upstream
or downstream plasma rest frame respectively. We will use downstream $r_{g}$
as a distance and $r_{g}/c$ as a time units. The magnetic field inclination
to the shock normal upstream of the shock, $\psi$, is measured in the
upstream plasma rest frame.

Let us denote the ratio of the cross-field diffusion coefficient
$\kappa_\perp$ to the parallel diffusion coefficient $\kappa_\|$ as $\tau$
(the value is measured in the plasma rest frame). Simulations prove that
fluctuations upstream of the shock (measured by $\tau^{U}$) and downstream
of the shock (measured by $\tau^{D}$) influence the acceleration process
independently. The minimum fluctuations upstream of the shock needed to run
the acceleration process efficiently tend to zero when
$\gamma \rightarrow \infty$. We checked by simulations with different
$\tau^{D}$ that its value does not influence the spectral
index considerably for a given $\tau^{U}$.

Our scattering model is very simple but also universal. In the model we
are not able to discuss gyroresonat scattering. Upstream of the shock
particles have not enough time to interact resonantly with low-frequency
waves and the rough relation $\tau \sim (\delta B/B)^{4}$
(cf. Blandford \& Eichler 1987) cannot be deduce from the
interaction there.
However, for growing $\tau^{U}$ and fixed $\Delta \vartheta$ the time
between scattering acts decreases what is equivalent to increasing the
magnetic field fluctuations.

\section{Results}

In the following simulations we consider shocks with $\gamma=$
20, 40, 80, 160, 320, magnetic field inclinations $\psi=15^\circ, 30^\circ,
45^\circ$, $60^\circ, 75^\circ, 90^\circ$ and downstream values of magnetic
field fluctuations
$\tau^{D}=0, 1.0\cdot 10^{-3}, 1.1\cdot 10^{-2}, 0.11, 0.69$.

Thus, as a first case we consider downstream conditions without magnetic field
fluctuations. By simple data inspection (cf. Fig.~2) we look for minimum
$\tau^{U}$ where the spectral index reaches its limit of 2.2 and we apply
this value in further simulations. The relation between $\tau^{U}$,
$\gamma$ and $\psi$ can be roughly fitted with the equation
$\tau^{U} = 0.25\, \gamma^{-1.2}\, \psi$ in the considered range of shock
parameters. We repeated simulations for a number of cases with different
$\gamma$ and $\psi$ and $\tau^{D}\not = 0$. The obtained results are in
good agreement with the ones derived from the above equation up
to $\tau^{D} = 0.11$.

Values of the acceleration time $t_{acc}$ for three amplitudes of magnetic
field fluctuations downstream of the shock are presented in Fig.~3.
In the figure one can see the lack of change of $t_{acc}$ with $\psi$, but
it slowly decreases to the asymptotic value with $\gamma$. In the simulations
we have observed tendency of $t_{acc}$ to grow when $\sigma$ increases up
to 2.3-2.4 and no further change if magnetic field fluctuations upstream of
the shock grow. For $\tau^{D}\leq0.11$ the asymptotic value of the
acceleration time is close to $r_{g}/c$. It occurs that $r_{g}/c$ is a good
unit provided that the homogeneous magnetic field dominates the randomly
component. Unfortunately, when this condition fails the meaning of $t_{acc}$
becomes unclear in the simulations then. For this reason we will not discuss
further the case of $\tau^{D} = 0.69$ any more.

Approximate calculations of Gallant \& Achterberg (1999) showed that
$t^{U}_{U}/t^{D}_{U}\simeq 1$, where $t^{U}_{U}$ is the particle mean residence
time upstream of the shock (upper index) as measured in the upstream plasma
rest frame (lower index), and D in $t^{D}_{U}$ stands for the downstream
residence time. However, they were not able to consider the anisotropic
particle momentum distribution and our results in Fig.~4 {\it transformed to
the upstream plasma rest frame} with $t^{U}_{U}/t^{D}_{U}$ within the range
$0.01-0.1$ are more adequate for real situations. Additionally, the above
authors applied an extremely irregular magnetic field upstream of the shock
represented by randomly oriented magnetic cells with field amplitude $B$ and
they measured time in the upstream unit of $r_{g}(B)/c$. As a result they
obtained that $t^{U}_{U}/t^{D}_{U}$ could be much larger than 1 in the case.

Just before the spectral index reaches its minimal value (cf. Fig.~2)
$\Delta \Omega_{S}$ stabilizes near the limit which value does not further
depend on the magnetic field inclination as is seen in Fig.~5. Momentum
vectors of particles crossing downstream of the shock have similar
distributions as measured in the downstream plasma rest frame if
$\Delta \Omega_{S}$ approaches the maximum value. Then, it follows that
parameters we consider below depend only on $\tau^{D}$.

\begin{figure}
\vspace*{11.8cm}
\includegraphics{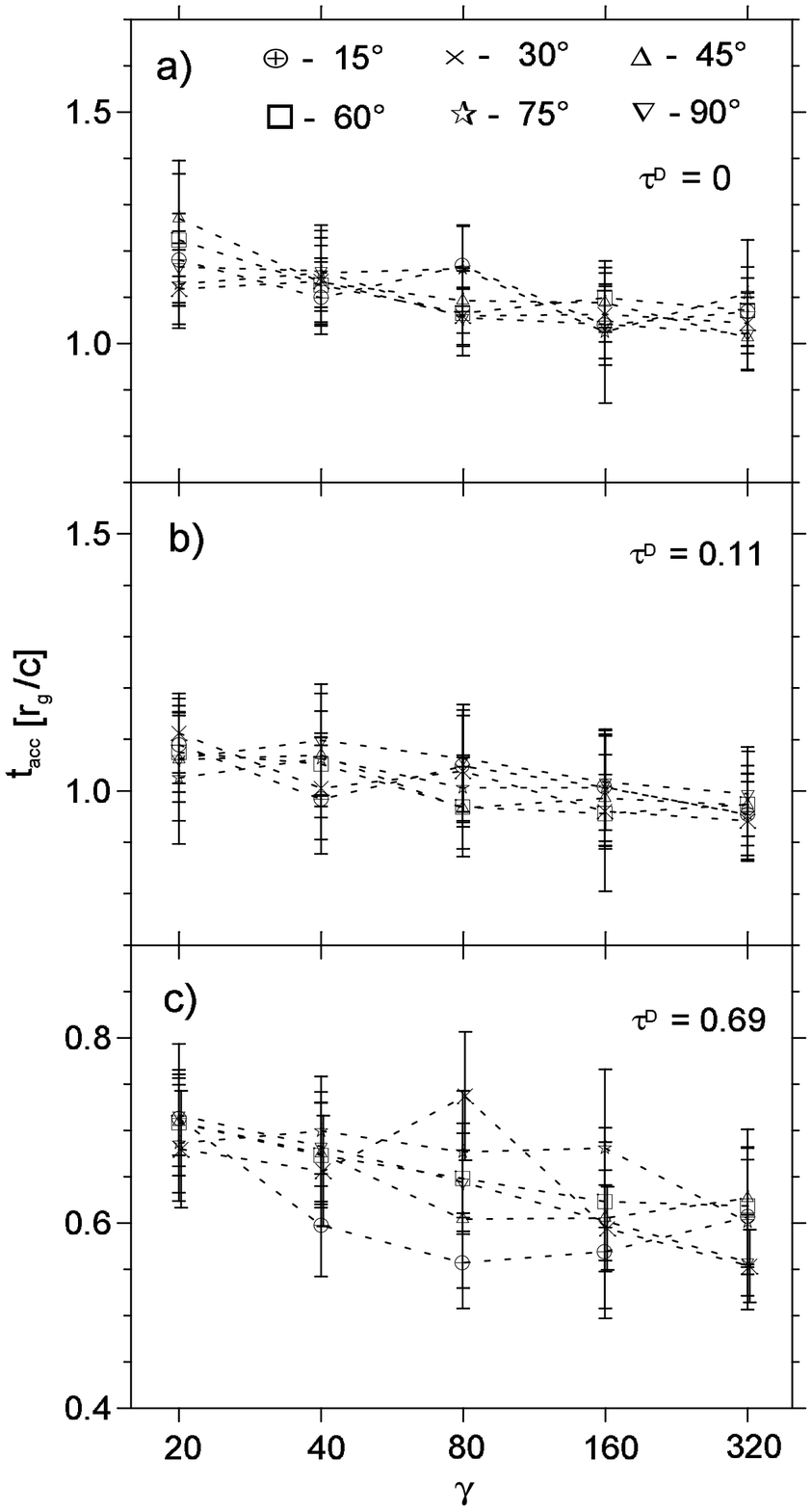}
\caption{
The simulated acceleration time as a function of the shock Lorentz factor:
a) without fluctuations downstream of the shock, b) with fluctuations downstream
of the shock, c) fluctuations downstream of the shock dominate homogeneous
magnetic field. Results for a given upstream magnetic field inclinations, given
in a panel a), are joined with dashed lines.}
\label{fig3}
\end{figure}

For growing $\tau^{D}$ ($\tau^{D} = 0,\, 1.0\cdot 10^{-3},\, 1.1\cdot 10^{-2},\, 0.11$)
\footnote{Below, we provide the respective series of simulated parameter for this
sequence of $\tau^{D}$} the acceleration time is constant and accompanied by
a slow increase of the mean energy gain in one cycle downstream-upstream-downstream
$\langle\Delta E/E\rangle_{D} = 0.89,\, 0.94,\, 1.0,\, 1.1$, and a slight
decrease of the fraction of particles that reach the shock again after crossing
it downstream $\langle\Delta n/n\rangle = 0.51,\, 0.50,\, 0.48,\, 0.44$.
Simultaneously the mean time a particle spends downstream of the shock grows as
$t^{D}_{D} = 0.96,\, 1.0,\, 1.2,\, 1.35$. Time that a particle spends upstream
of the shock can be neglected in this rest frame as is visible in Fig.~4. It
implies, approximately, $t_{acc} = t_{D}^{D}/\langle\Delta E/E\rangle_{D}$ if one
neglects correlations between these quantities (cf. Bednarz \& Ostrowski 1996).
Similarly we can roughly estimate the value of the energy spectral index of
accelerated particles as
$\sigma\simeq 1- \ln(\langle\Delta n/n\rangle)/\ln(\langle\Delta E/E\rangle_{D}+1)$.

\begin{figure}
\vspace*{5.8cm}
\includegraphics{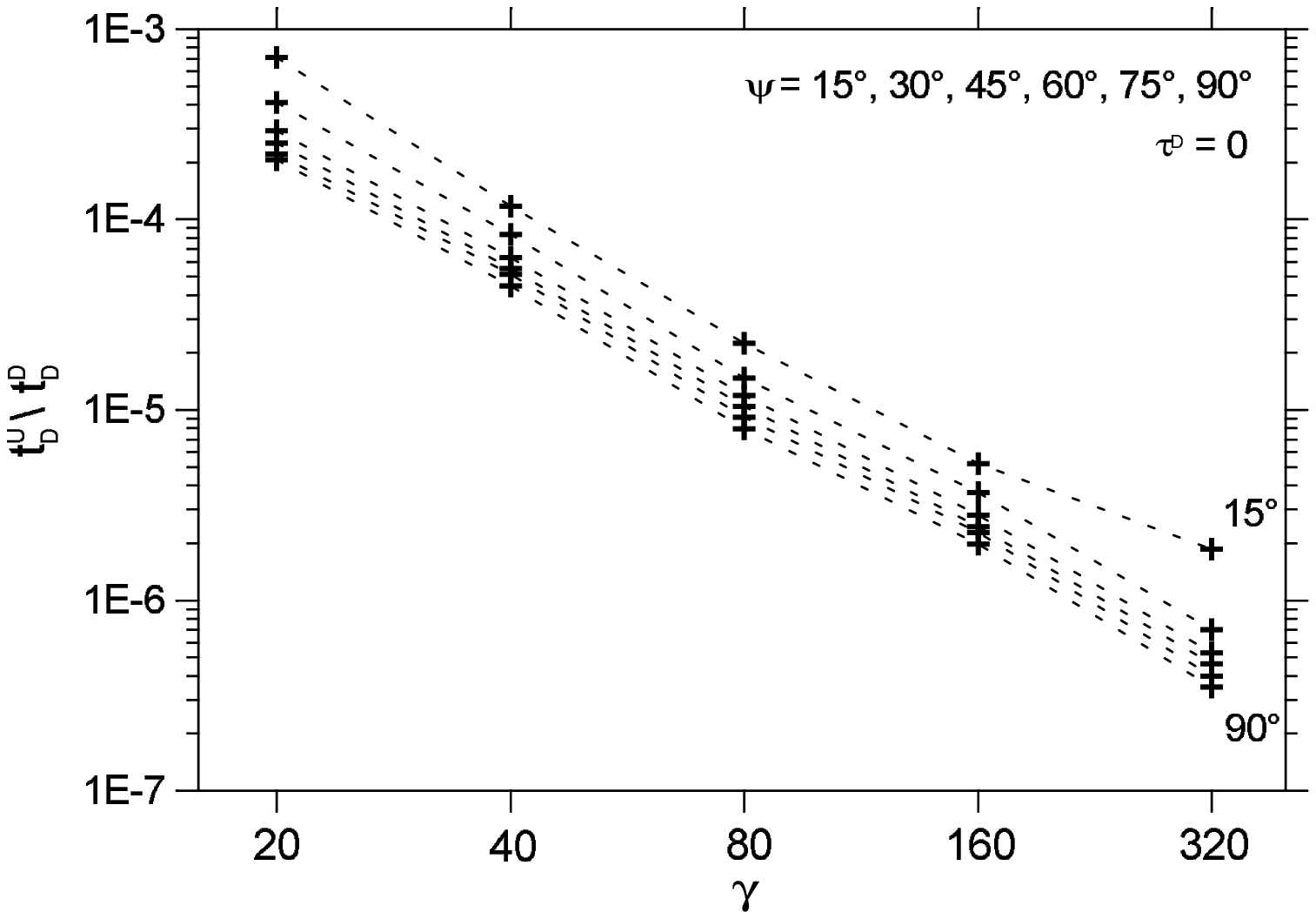}
\caption{
The ratio of the mean time a particle spends upstream of the shock between shock
crossings to the time
it spends downstream of the shock as a function of the shock Lorentz factor.
It slightly decreases with growing upstream magnetic field inclination. Dashed
lines join points with a constant $\psi$. Apparent deviation of the point with
$\gamma=320$, $\psi=15^\circ$ is real.}
\label{fig4}
\end{figure}

The simulated maximum distance in downstream medium the particle is able to
depart from the shock and reach it again is, respectively,
$d^{M}_{D}\, =\, 0.84,\, 1.5,\, 2.5,\, 4.0$
(the values were derived from $\sim 10^{5}$ events).
We calculated the average values of $\sin^{2}\theta$ for returning particles
wandering downstream of the shock and found, respectively,
$\langle\sin^{2}\theta\rangle = 0.676,\, 0.658,\, 0.650,\, 0.651$.

\begin{figure}
\vspace*{5.8cm}
\includegraphics{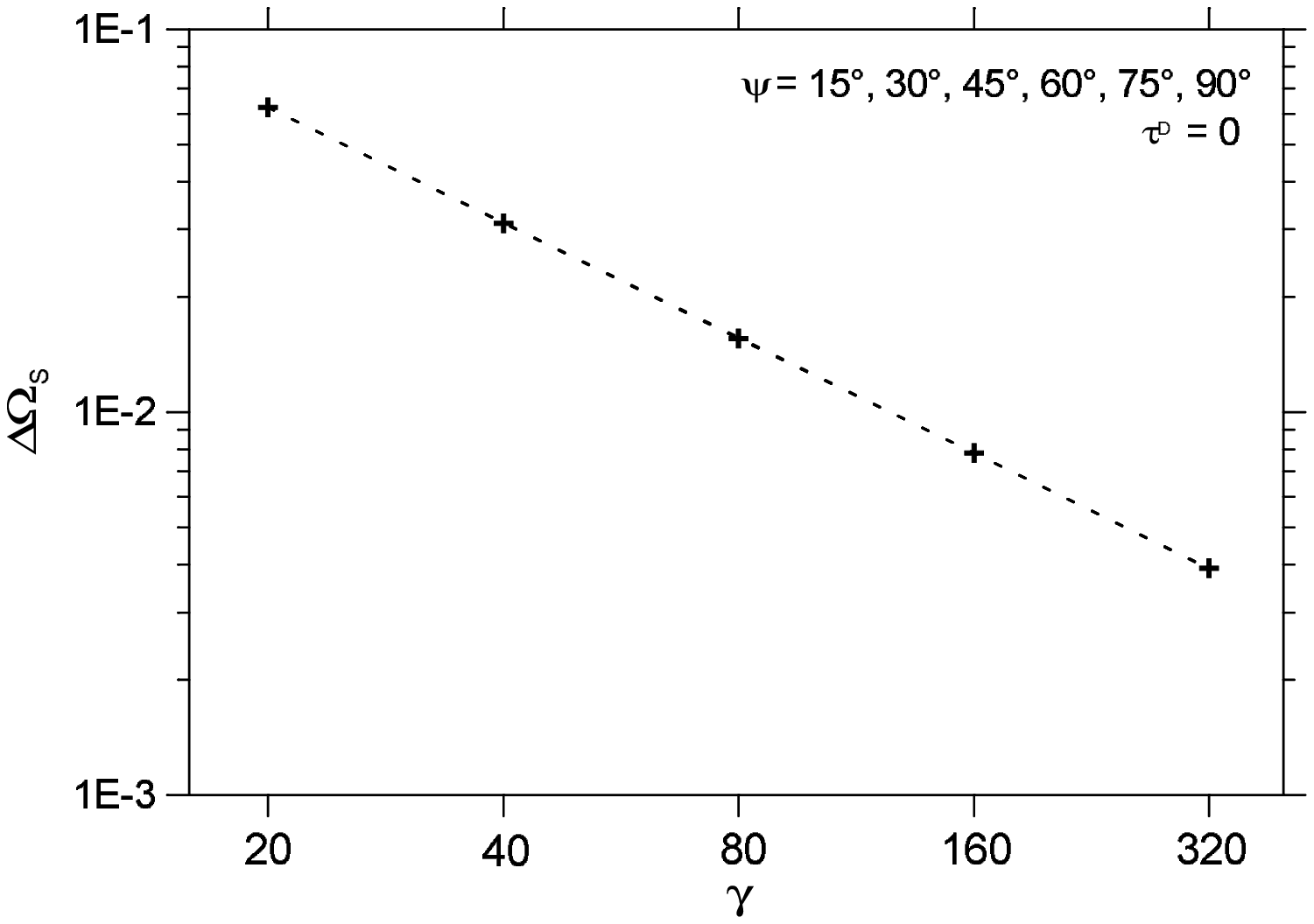}
\caption{
The mean deflection angle resulting from scattering in an inhomogeneous magnetic
field upstream of the shock as a function of the shock Lorentz factor
$\gamma$ with $\tau^{U}$ chosen in a way as illustrated in Fig.~2. Six dashed
lines for different $\psi$ are identical.}
\label{fig5}
\end{figure}

\section{Discussion}
Because of newly found acceleration mechanism in ultrarelativistic
shock waves we propose that some part of GRBs radiation could arise
due to synchrotron radiation of electrons or electron pairs accelerated
across the mechanism. We follow the internal shocks model of GRBs
(cf. Kobayashi et al. 1997 for example). In the model two different shells
have different Lorentz factors. The inner shell overtakes a slower outer
shell and form a shock. The Lorentz factor of the shock as measured in the
frame at rest with respect to the outer shell is assumed to be $\sim 2$.
With only a part of kinetic energy converted into the internal energy the
particle energy distribution downstream of the shock will be non-thermal
with, possibly, a substantial fraction of relativistic particles.
As a result an amount of relativistic particles will be present in the
shock. The particles can be accelerated across the mechanism presented in
the paper and could be observed in the afterglow (cf. Waxman 1997,
Galama et al. 1998).

We derived some parameters of the process that could be used in
GRBs models. The acceleration time $t_{acc} = 1.0\, r_{g}/c$, measured
in the downstream plasma rest frame in the unit of particle gyroradius in
the homogeneous magnetic field component divided by the speed of light is
the second important parameter besides the spectral index $\sigma = 2.2$.
The values of $d^{M}_{D}$ and $t^{D}_{D}$ define the dimensions of the
shock that allow the process to be effective, and the values of
$\langle\sin^{2}\theta \rangle$ shows how the synchrotron radiation can
influence the process.

In the ejecta of the relativistic matter in the GRB model the outer shells
can be faster than the following ones. In this case separated shocks with
Lorentz factors reaching $\gamma \sim10^{3}$ will be generated. The leading
shock could produce seed protons with energies of $10^{14}-10^{16}$ eV.
These protons downstream of the first shock can interact with the following
one to be reflected with energy gains $\sim \gamma^{2}$ (cf. Gallant \&
Achterberg 1999, Bednarz \& Ostrowski 1999). For a constant reflection
probability the spectrum of these reflected highest energy particles,
above $10^{20}$ eV, will be only the shifted in energy spectrum
of seed particles with the universal spectral index $\sim2.2$.

\section*{Acknowledgments}

The author is grateful to Micha{\l} Ostrowski for valuable discussions
and the referee for comments that was helpful in clarifying the contents
of this paper. The
presented computations were partly done on the HP Exemplar S2000 in
ACK `CYFRONET' in Krak\'ow. The present work was supported by
{\it Komitet Bada\'n Naukowych} through the grant PB~179/P03/96/11
and PB~258/P03/99/17.

\end{document}